\documentclass[12pt]{article}

\begin{document}

\title {Phenomenology of magnetically active superconductors}
\author {by\\DOMINIK ROGULA\thanks{Polish Academy of Sciences,
Institute for Fundamental Technological Research,
ul. Pawi\a'{n}skiego 5B, PL-00-106 Warsaw, E--mail: dominikrogula@op.pl
}
 and MA{\L}GORZATA SZTYREN
\thanks{Department of Mathematics and Information Science,
Warsaw University of Technology, Pl. Politechniki 1, PL-00-661 Warsaw,
E--mail: emes@mech.pw.edu.pl}}
\maketitle

\newcommand{\equn}[1]{\begin{equation} {#1} \end{equation}}
\newcommand{\equln}[2]{\begin{equation} {#1} \label{#2} \end{equation}}

{\footnotesize {\bf Summary:}
   The thermodynamical phenomenological theory of magnetically active 
superconducting materials and magnetic-superconducting heterostructures
is presented. The materials may exhibit arbitrarily strong anisotropy,
parametric or structural. Exact anisotropic similarity transformations
are found for both the continuum and layered hybrid systems.

\normalsize
\section{Introduction}
   A formulation of thermodynamical theory of magnetically active,
anisotropic materials admitting coexistence of the superconductivity
and magnetic order has been proposed in \cite{Rogula 2004}. The
theory is based on the
Ginzburg-Landau approach extended to multi-component order parameters
and the states of thermodynamic
quasi-equilibrium far from the superconducting phase transition.
The field equations are derived under assumption of the \(U(1)\)
gauge invariance.
The nonlinearity with
respect to the order parameter, magnetic field, and other physical
quantities is assumed reasonably general, without any particular
restrictions of its functional form.

   In the present paper we give an improved version of that theory
complemented with the exact anisotropic similarity
transformations for layered hybrid systems \cite{Sztyren:2006a}.

\section{Background}
   A material medium is considered magnetically active if
its magnetic properties  differ (substantially) from those of the
physical vacuum. Although the superconductivity itself is a magnetic
phenomenon, the superconducting materials need not be magnetically
active. This circumstance justifies disregarding the magetic activity
of a superconducting material in heuristic or didactic considerations,
but not in a more complete theory. Due to orbital polarizability and/or
spin degrees of freedom, the effects of magnetic activity of
superconducing materials are practically always present, and in many
instances significant. In particular it concerns the heterostructures
which involve interactions between superconducting and ferromagnetic
components \cite{Bourgeois+ 2001,Stefanikis+Melin 2003}.
One may also notice that,
on the background of some controversial views
\cite{Berk+Schrieffer 66, Ginzburg 57} concerning coexistence
of superconductivity and ferromagnetism, materials which combine
these phenomena were found experimentally \cite{Aoki+ 2001,
Huxley+ 2001, Pfeiderer+ 2001, Saxena+ 2000}.
Moreover, it seems that, in accordance with earlier theoretical
predictions, the superconductivity of such materials
is magnetically mediated \cite{Karchev 2003}.

   In order to describe the superconducting state of condensed matter,
the {\sc Ginzburg-Landau} theory \cite{Ginzburg+Landau} introduces
a phenomenological quantity, called the order parameter, represented
by a complex scalar field \(\psi(x)\) with properties quite
analogous to the microscopic quantum-mechanical single-particle
wave function. Contrary, however, to the microscopic wave function,
the GL order parameter describes the macroscopic condensate of the
supercurrent carriers and enters as an independent variable into
the thermodynamical free energy functional.
 This approach turned out to be extremely succesful.

   In the original GL paper the functional was given a simplest form
suitable for description of the states of a superconducting material
near the phase transition. In this form, the theory was confirmed
afterwards by {\sc Gorkov} \cite{Gorkov 59a} through rederivation
from the microscopic theory. Its range of applicability
turned out, however,
severely restricted by the condition of proximity to the
transition point.

   To cope with broader ranges of real situations, the GL theory
has been extended in various directions, including multicomponent
order parameters, more elaborate forms of the free energy
functional, time-dependent phenomena, and heterostructures.
 Any such modification,
as long as it respects the idea of a (possibly generalized)
superconducting order parameter as a thermodynamical variable,
is referred to as belonging to
the framework of {\em extended} Ginzburg-Landau phenomenology
\cite{Rogula 99,Rogula+Sztyren:2006,Sztyren:2006a}.

  The theory under consideration admits
arbitrary (in a reasonable sense) functional forms of the GL
thermodynamical potential. We intend to include in this way also
the thermodynamical states which are far from the superconducting
phase transition.
We admit multiple order parameters, anisotropy,
and material inhomogeneities, but,
for simplicity, we retain the assumptions of stationarity,
thermodynamical quasi-equilibrium, and
charge neutrality.

   The assumption of charge neutrality refers to the resultant
electric charge density of all charged components, such as
the superconducting carriers (Cooper pairs), normal electrons,
and the ionic substrate.
It restricts our considerations to the case of bulk
superconductivity in good
(quasi-metallic) superconductors. To include "bad" superconductors
and interface effects
one would need a further extension of the model, which is beyond the
scope of the present paper and will be given elsewhere.

    We shall focus our attention on
superconducting materials which can be (a) strongly
anisotropic, (b) inhomogeneous at the macro- and micro-scale,
and (c) magnetically active. Below we shall comment briefly on
the above marked specific assumptions.

  (a) The idea of isotropic superconductor, due to its simplicity
and heuristic power, attracts much attention as
a tool in theoretical investigations. It suffices here to
note examples of most outstanding and successful classic theories
of superconductivity such as Ginzburg-Landau, 
Bardeen-Cooper-Schriefer
, and Abrikosov's   
 ones, each of them based on an appropriate
isotropic model. On the other hand, the real superconducting
materials exhibit, as a rule, 
some anisotropy of physical
properties.
In particular, the anisotropy of high \(T_c\) cuprate oxides is
rather high
\cite{
Blatter+ 94,Cyrot+Pavuna}: e.g. the ratio of penetration depths
\(\lambda_c/\lambda_{ab}\) for YBa\(_2\)Cu\(_3\)O\(_{7-y}\)
and Bi\(_2\)Sr\(_2\)CaCu\(_2\)O\(_{8+y}\) lies in the range
\(5{\div}7\) and \(50{\div}200\), respectively; it corresponds to estimated
values
of the ratio of flux line tensions \(U_c/U_{ab}\) ranging from 20
to 40000.

  (b) To describe heterogeneous superconducting materials or
devices, the theory must admit material inhomogeneities of various
-- macroscopic, microscopic, and intermediate -- scales of length.
The inhomogeneities can be smooth or abrupt, technological or
natural.
Many high T\(_c\) superconducting materials exhibit a distinct
layered structure at the atomistic level. The spacings
between atomic layers, such as Cu-O
 planes in cuprate-oxide
superconductors, can be relatively large as compared with the
in-plane distances what, besides the structural anisoptropy,
can result in a natural micro-scale inhomogeneity of those materials
in the out-of-plane direction. The {\sc Lawrence-Doniach} model
\cite{Lawrence+Doniach 71} and its generalization
\cite{Rogula+Sztyren:2006,Sztyren:2006a} can be considered an extreme
case of such a natural micro-inhomogeneity.
On the other hand, due to the recent progress in nanotechnology
it is possible to fabricate diverse nano-scale technological
heterostructures.

   (c) Apart from magnetic activity
originating from physical factors,
we found it also expedient to introduce its equivalent
at the level of theoretical modelling.
We refer here to the anisotropic similarity transformations -
a useful
tool in the study of anisotropic superconductors, discussed in
Section \ref{Secani}.

\section{Notation}\label{notation}
    Whenever appropriate, we use implicit
tensor notation with supressed indices; the indices are exposed
only when otherwise the expressions would be equivocal or illegible.
For instance,
a point in the physical space (as well as the corresponding material
point when the distinction makes no difference) is
denoted simply by \(x\),
which is to be developed to \((x^1,x^2,x^3)\) in typical 3D
situation.

    Basically, the standard notation
from recent literature on phenomenology of superconductivity
\cite{Ketterson+Song} is employed throughout the paper.
The effective mass \(m^*\) of the supercurrent carriers is tensorial;
in the case of transversally isotropic
superconducting materials
(e.g. layered oxides, in good approximation) the relevant
mass tensor components are \(m_{ab}\) (in-plane) and
\(m_c\) (out-of-plane).

    The complex conjugate of the order parameter \(\psi\)
is denoted by \(\bar\psi\). The symbols \(\partial\) and
\(\nabla\) are employed in a somewhat non-standard way.
The partial derivative operator is
denoted by \(\partial\), while the symbol \(\nabla\) is reserved for
the electromagnetic \(U(1)\)-covariant derivative.
Both operators \(\partial\) and \(\nabla\) act from the left.
We found it expedient to introduce also the conjugate operators
\(\bar\partial\) and \(\bar\nabla\) which act from the right.

    Due to a bug in the routine translating from TeX to PDF, in the
final printed form of the reference \cite{Rogula 2004} a repeated
typographic error occured. The error consisted in the fusion of two
consecutive short overbars into a longer one. It resulted
in erroneous replacement of the
expressions
\(\bar\psi\,\bar\partial\) with \(\overline{\psi\partial}\) and
\(\bar\psi\,\bar\nabla\) with \(\overline{\psi\nabla}\).
The present text contains terms counteracing those faults.

\section{The reference model}\label{refer}
    To fix the point of departure for construction of extended models
we start from the reference model which, excepting the details
of notation, almost exactly coincides with the original GL model
\cite{Ginzburg+Landau}. It is also convenient to develop the notation
adapted in this paper.

   1. The superconducting state of the superconductor is described
with the aid of the order parameter represented by a single-component complex
scalar function \(\psi=\psi(x)\).

    The gauge transformation for this parameter is defined as
\equln{\left\{\begin{array}{l}
   A \rightarrow A+\partial\Lambda,\\
   \psi\rightarrow e^{\frac{ie^{*}}{\hbar c}\Lambda}\psi,\\
   \bar\psi\rightarrow \bar\psi e^{\frac{-ie^{*}}{\hbar c}\Lambda},
   \end{array}\right .}{gauge}
where \(e^*\) denotes the electric charge of the supercurrent carriers.
The corresponding covariant derivatives \(\nabla\) and \(\bar\nabla\)
are given by
\equln{\nabla=\partial-\frac{ie^*}{\hbar c}A,}{nab}
and
\equln{\bar\nabla=\bar\partial+\frac{ie^*}{\hbar c}A.}{nabc}

   2. The free energy of the superconductor is given by the
functional
\equln{F[T;\psi, A]=\int d^3x F_S}{locF}
with an appropriate free energy density \(F_S\) of the superconducting
state.

    3. The free energy density \(F_S\) equals
\equln{F_S=F_N+\alpha\mid\psi\mid^2
       +\frac{\beta}{2}\mid\psi\mid^4
       +\frac{\hbar^2}{2m^{*}}
         \bar\psi\,\bar\nabla\nabla\psi,}
{GLF}
where \(F_N\) refers to the normal state and
\(m^{*}\) represents the effective mass of the supercurrent carriers.
The coefficients \(\alpha\) and \(\beta\) are given as
\equln{\alpha=a(T-T_c),\ \ \ a=const,\ \beta=const.}{albet}
The normal state free energy density \(F_N\)
does not depend on the order parameter.

The expression (\ref{GLF}) ensures the already required invariance
with respect to the gauge transformations (\ref{gauge}).

\section{Local first order functional}\label{firsto}
   The reference model sketched in the previous section relies upon
many intentional simplifications, including isotropy, homogeneity,
and proximity to the N-S transition point. Some of these drawbacks
are direct consequences of the particular form of the free energy
density \(F_S\) and can be eliminated simply by admitting,
in place of (\ref{GLF}),
more general functionals defined on the same space of the field
variables. In this way one can successfully take into account many
realistic features of superconductors. As examples one can mention
such properties as high anisotropy
or more subtle thermodynamical behaviour of high \(T_c\)
superconductors \cite{Shehata 89,Blatter+ 94,Ketterson+Song}.

   The range of applicability of such models is, however, severely
restricted by the singlet nature of the order parameter.
From the microscopic point of view a singlet order parameter
is justified, first of all, for superconductors with dominating
isotropic \(s\)-pairing. Also other types of pairing
can be treated in this way, provided that a single pairing
mode, corresponding to a one-dimensional representation of the
internal symmetry group, is strongly dominating.

   Many superconding materials require, however, multiple order
parameters.  One can mention here the heavy fermion unconventional
superconductors \cite{Sigrist+Ueda 90} which exhibit anisotropic
spin triplet \(p\)-pairing, possibly mixed with spin singlet
\(s\)-pairing. Another typical example is delivered by the coexistence
of spin singlet \(s\)- and \(d\)-pairings in orthorombic
high \(T_c\) materials \cite{Walker 96}.

   In the present section we shall briefly discuss an extension of
the GL phenomenology which avoids the above mentioned
drawbacks.

   1. First of all we modify the assumption stated in item 1
of Section 1. In place of the singlet order parameter, represented
in the reference model by
a complex-valued scalar field, a more general
multi-component order parameter is introduced. For brevity,
this order parameter will also be denoted by the single symbol
\(\psi=\psi(x)\). In geometrical interpretation the admissible
values of \(\psi\) will be considered as organized into a differential
manifold. Whenever a particular component will be referred to,
an indexed symbol like \(\psi^A=\psi^A(x)\) will be used and interpreted
as an appropriate -- real or complex -- coordinate in the corresponding
local map. The index \(A\) will take integer values from 1 to \(N\),
with arbitrary fixed \(N\). The real dimension of the order parameter
manifold may, due to the presence of complex-valued coordinates,
differ from \(N\).

   2. From the physical point of view the multiplet order parameter
represents generally a mixture of charged fields. In contrast
to the reference model, such a mixture can not be characterized by
a single scalar charge. To characterize the coupling of the
multiplet order parameter \(\psi\) with the electromagnetic field
we must specify an adequate gauge transformation. To that end,
in place of the scalar charge \(e^*\)
we introduce a real symmetric matrix
\equln{e^*=(e^{*}_{\ AB})}{matcharge}
such that the gauge transformation is defined by the corresponding
matrix reinterpretation of the old formula (\ref{gauge}). In this way
also the gauge covariant derivatives (\ref{nab}) and (\ref{nabc})
retain their graphic form with reinterpreted meaning.

   In general, the charge matrix \(e^*\) need not be diagonal.
However, being real and symmetric, it can be diagonalized by an
appropriate orthogonal transformation of the order parameters
\(\psi^A\). The eigenvalue 0, if present, corresponds to
electrically neutral subspace of order parameters which
carries no supercurrent; in spite of this property such order
parameters can seriously
affect the thermodynamical properties of a superconducting system.
The remaining eigenvalues of \(e^*\) determine the spectrum
of charges of supercurrent carriers.

   In most cases the supercurrent carriers are experimentally
identified as electrically equivalent to
pairs of electrons or holes. Then the spectrum of the
matrix \(e^*\) simplifies to at most 3 eigenvalues: \({\pm}2e\) and 0.

   3. We accept the assumption formulated in item 2 of Section
\ref{refer}, so that the free energy functional retains the local
form (\ref{locF}).
We, however, modify item 3.: instead of the particular form
(\ref{GLF}) we assume
%
\equln{F_S=F[\psi, A]=F(T, x, \psi, \bar\psi, \partial\psi,
\bar\psi\,\bar\partial, A, \partial A).}{fiord}
The explicit dependence on \(x\) is intended to reflect
natural or artificial
inhomogeneities of the materials, such as inhomogeneous doping,
structure defects, etc., including possible
heterostructural properties of the system.
Implicit dependence on some external control
parameters (such as mechanical deformation) is admitted; the
implicit parameters can also depend on the position \(x\).

   Apart from reasonable mathematical properties and the gauge
invariance, no particular restrictions will, in general,
be imposed upon the function (\ref{fiord}). The symbol \(e^*\)
stands here for the matrix (\ref{matcharge}).

   4. The requirement of gauge invariance results in the following
relations
\equln{\frac{\partial F}{\partial\psi}e^*\psi-
       \bar\psi{e^*}\frac{\partial F}{\partial\bar\psi}+
       \frac{\partial F}{\partial(\partial\psi)}e^*\partial\psi-
       \bar\psi\,\bar\partial{e^*}
       \frac{\partial F}{\partial(\bar\psi\,\bar\partial)}
       =0,}{ginv1}
\equln{\frac{\partial F}{\partial A}+\frac{i}{\hbar c}
       (\frac{\partial F}{\partial(\partial\psi)}e^*\psi-
        \bar\psi{e^*}\frac{\partial F}
        {\partial(\bar\psi\,\bar\partial)})
       =0,}{ginv2}
and
\equln{\frac{\partial F}{\partial(\partial_j A_k)}+
       \frac{\partial F}{\partial(\partial_k A_j)}
      =0}{ginv3}
to be satisfied identically by the function (\ref{fiord}).

   The conditions (\ref{ginv2}) and (\ref{ginv3}) can be integrated
in a straightforward way. As a result, one obtains
the functional dependence
\equln{F_S=F[\psi, A]=F(T, x, \psi, \bar\psi, \nabla\psi,
\bar\psi\,\bar\nabla, B)}{figi}
with the vector potential \(A\) entering indirectly through the
gauge covariant derivatives \(\nabla, \bar\nabla\) and the
(gauge invariant) magnetic induction field \(B={\rm curl}\,A\).
The gauge invariance conditions for the function (\ref{figi})
reduce to
\equln{\frac{\partial F}{\partial\psi}e^*\psi-
       \bar\psi{e^*}\frac{\partial F}{\partial\bar\psi}+
       \frac{\partial F}{\partial(\nabla\psi)}e^*\nabla\psi-
       \bar\psi\,\bar\nabla{e^*}\frac{\partial F}
       {\partial(\bar\psi\,\bar\nabla)}
       =0.}{ginv1a}

    5. Taking into account the functional dependence (\ref{figi}),
the functional derivatives of \(F_S\) with respect to the field
variations \(\delta\psi\), \(\delta\bar\psi\), and \(\delta A\)
can be represented in the form
\equln{\frac{\delta F}{\delta\psi}=\frac{\partial F}{\partial\psi}-
   \nabla\frac{\partial F}{\partial(\nabla\psi)},}{eqpsi}
\equln{\frac{\delta F}{\delta\bar\psi}=
       \frac{\partial F}{\partial\bar\psi}-
   \frac{\partial F}{\partial(\bar\psi\,\bar\nabla)}\bar\nabla,}{eqbar}
and
\equln{\frac{\delta F}{\delta A}=\frac{i}{\hbar c}
    (\bar\psi{e^* }\frac{\partial F}{\partial(\bar\psi\,\bar\nabla)}-
      \frac{\partial F}{\partial(\nabla\psi)}{e^*}\psi
    )+ {\rm curl}\ \frac{\partial F}{\partial B}
  .}{eqA}

   6. In consequence, one obtains the generalized Ginzburg-Landau
equation
\equln{\frac{\partial F}{\partial\bar\psi}-
   \frac{\partial F}{\partial(\bar\psi\,\bar\nabla)}\bar\nabla=0,}
   {eqGL}
and the magnetic field equation
\equln{{\rm curl}\, H=\frac{4\pi}{c}\,j}{eqH}
with the intensity of magnetic field
\equln{H=4\pi\,\frac{\partial F}{\partial B}}{defH}
and the supercurrent density
\equln{j=-\frac{i}{\hbar}
    (\bar\psi{e^* }\frac{\partial F}{\partial(\bar\psi\,\bar\nabla)}-
      \frac{\partial F}{\partial(\nabla\psi)}{e^*}\psi
    ).}{defj}

   7. Taken literally, the above equations are derived for sufficiently
smooth fields. As a rule, such degree of differetiability can not be
guaranteed at abrupt interfaces, in particular at the heterostructural
ones. However, the integral form of those equations is valid
under much weaker differentiability conditions, sufficient for
the derivation of boundary conditions. The standard procedure applied
to the equation (\ref{eqpsi}) leads to the boundary conditions, which
in index notation take the form of continuity equation across the
interface
\equln{\frac{\partial F^I}{\partial(\bar\psi^A\bar\nabla_k)}n^k=
\frac{\partial F^{II}}{\partial(\bar\psi^A\bar\nabla_k)}n^k};
where \(n^k\) represents the normal vector of the surface
demarcating the regions I and II. In consequence of this equation
the normal component of the supercurrent (\ref{defj}) is
continuous.

\section{Layered hybrid model}
The free energy functional for higher-grade hybrid model
\cite{Sztyren:2006a} has the following form 
\begin{equation}
{\cal{F}}={\cal{F}}_0+{\cal{F}}_S+\int g({\bf B})d^3x.
\label{calf}
\end{equation}
The last term represents the contribution of magnetic field to
the free energy. 
In the standard particular case of magnetically inactive material the 
function $g({\bf B})$ reduces to ${\bf B}^2/8\pi$.
The term \({\cal F}_0\) describes the normal state, while \({\cal F}_S\)
the superconducting one. The supercoducting term is composed
of two parts: 
\begin{equation}
{\cal{F}}_S={\cal{F}}_P+{\cal{F}}_J,\label{fs}
\end{equation}
where the part
\begin{equation}
{\cal{F}}_P=s\sum_n\int dxdyF_n
\label{efn}
\end{equation}
describes the contribution of atomic planes, while
\({\cal{F}}_J\) corresponds to interplanar Josephson's bonds.
\(F_n\) has the Ginzburg-Landau form (\ref{GLF}), with 2D operator
$\nabla$ and the coefficients dependent, in general, on the plane \(n\).

The form of the functional \({\cal{F}}_P\) already 
ensures its invariance with respect to the gauge transformation
(\ref{gauge}), understood this time in the 2D sense.

The Josephson term  in the free energy functional for the hybrid
model of grade K has the form
\begin{equation}
{\cal{F}}_J=s\sum_{n}\sum _{q}
\int dxdy\,\varepsilon_{qn}.\label{fj}
\end{equation}
The quantity \(\varepsilon_{qn}\) denotes the energy
of J-link
between n-th and (n+q)-th planes and is given by
\begin{equation}
\varepsilon_{qn}=\frac{1}{2}\{\zeta_q
(|\psi_n|^2+|\psi_{n+q}|^2)
-\gamma_q
 (\bar{\psi}_n\psi_{n+q}e^{-ip_{qn}}+c.c.)\},\label{epsi}
\end{equation}
with the coupling parameters \(\zeta_q\) and \(\gamma_q\) vanishing
for \(q>K\). 
The exponent \(p_{qn}\) is defined by the formula
\begin{equation}
p_{qn}=\frac{e^*}{\hbar c}\int_{ns}^{(n+q)s}A_zdz.\label{pqn}
\end{equation}

\section{Structured order parameters, nonlocal and higher
         order models}

    In the extension of GL phenomenology discussed in Section
\ref{firsto} we restricted our attention to superconductors
which can be described by local first order functionals
with multiplet order parameters. Such a restriction greatly
simplifies the theoretical considerations. On the other hand,
it eliminates
some physical phenomena which are known to be present in
superconductors. It concerns, in particular, the nonlocal
effects and nonparabolic dispersion curves.
Such phenomena can, however, be taken into account by
admitting yet more general -- nonlocal and/or higher order --
free energy functionals.

   The microscopic theories of superconducting materials of complex
atomic and electronic structure indicate various possible mechanisms
of superconductivity \cite{Anderson 97 book}.
To cope with the diversity
of those mechanisms, the phenomenological description can be
further generalized by introducing the structured order parameter.
Instead of the customary order parameter
represented by a (possibly multicomponent) function \(\psi(x)\),
we introduce a generalized one, represented by a function of
\(\psi(x, \xi)\), where the new variable \(\xi\) runs over
a certain internal space \(\Sigma\). The space \(\Sigma\)
serves as a mathematical stage on which the internal structure
of superconducting
carriers (such as Cooper pairs) appears.

   The space \(\Sigma\) depends on
the material under consideration and, parametrically, on the
temperature, doping, and other control parameters
such as external mechanical deformation. In the case of uniform
material samples the space \(\Sigma\) is the same for all
material points; in the case of heterostructures and inhomogeneous
material samples it can become \(x\)-dependent.

   The concept of structured order parameter offers tighter bound
between the phenomenology and the microscopic theories.
Apart from more accurate quantitative description it allows also
to study possible topological transitions relalated to
changes in the deep topological structure of the space \(\Sigma\).
The detailed discussion of those effects is, however, beyond the
scope of the present paper.

\section{Anisotropic similarity and scaling}\label{Secani}
   Assume now that the order parameter space is endowed with a
linear structure and consider a linear transformation
\(x \rightarrow x', \psi \rightarrow \psi', A \rightarrow A'\)
of the following form
\equln{\left\{ \begin{array}{l}
x'= Q x,\\
\psi'= R \psi,\\
A'= \tilde{Q} A,
\end{array}\right .}{anisim}
where \(Q\), \(R\), and \(\tilde{Q}\) are linear matrices;
the matrices are real except \(R\)
which can be, totally or partially, complex. The transformation rule
\equn{\partial' = \partial Q^{-1}}
is an immediate consequence of (\ref{anisim}).

  1. The transformation (\ref{anisim}) in its general form violates
the \(U(1)\) gauge
structure. If we request that, under (\ref{anisim}), the depart gauge
covariant derivative be transformed into the target gauge covariant
derivative,\(\nabla\psi \rightarrow \nabla'\psi'\), the transformation
(\ref{anisim}) must be adequately resticted. It follows that the relation
\(\bar{Q}=Q^{-1}\) must hold. Such a transformation will be called
a (gauge covariant) anisotropic similarity. The resulting transformation
rules for the gauge covariant derivatives of \(\psi(x)\) and
\(\bar\psi(x)\)  take the form
\equln{\nabla'\psi'=R Q^{-1T}\nabla\psi,\ \ \
       \bar\psi'\nabla'=\bar\psi\,\bar\nabla Q^{-1}R^+,}{covsim}
where the superscripts \(^T\) and \(^+\) stand for the matrix
transposition and Hermitian conjugation, respectively.

  2. Consider two material systems, say the dashed and
the undashed ones,
described by the thermodynamical functionals (\ref{locF})
with the densities \(F'(.)\) and \(F(.)\), respectively.
We shall say that those systems are anisotropically similar if
\equn{F'(T, x, \psi, \bar\psi, \partial\psi,
         \bar\psi\,\bar\partial, A, \partial A)=
         F(T, x', \psi', \bar\psi', \partial'\psi',
         \bar\psi'\bar\partial', A', \partial' A').}
Due to the gauge covariance this is equivalent to
\equln{F'(T, x, \psi, \bar\psi, \nabla\psi,
         \bar\psi\,\bar\nabla, B)=
         F(T, x', \psi', \bar\psi', \nabla'\psi',
         \bar\psi'\bar\nabla', B').}{Fdash}

   3.    To justify this conclusion let us examine the behaviour
of the field equations (\ref{eqGL} -- \ref{defj}) under the
trasformations (\ref{anisim}). Taking into account the
transformation rules (\ref{anisim} - \ref{covsim})
we obtain from eqn. (\ref{Fdash})
\equln{\frac{\partial F'}{\partial\bar\psi'}=
            R^{-1+}\frac{\partial F}{\partial\bar\psi},}{efdash}
\equn{\frac{\partial F'}{\partial(\bar\psi'\bar\nabla')}=
            R^{-1+}Q\,\frac{\partial F}{\partial(\bar\psi\bar\nabla)}.}
In consequence, from the expression (\ref{defj}) one obtains
\equn{j'=Q\,j.}
Further,
\equn{B'=({\rm det}\,Q)^{-1}Q\,B}
and
\equln{H'=({\rm det}\,Q)Q^{-1}H.}{Hdash}
The relations (\ref{efdash} -- \ref{Hdash}) guarantee the covariance
of the field equations (\ref{eqGL}) and (\ref{eqH}) with respect to
the anisoptropic similarity transformations.

  In the layered hybrid model a straightforward calculation shows
that the energy of the Josephson's coupling (\ref{epsi})
is invariant with respect to transformation
\begin{equation}
\left\{ \begin{array}{l}
\psi'_n=(-1)^n\psi_n,\\
{\bf A}'={\bf A},\\
\gamma'_q=(-1)^q\gamma_q,\\
\end{array}
\right .
\label{trans}
\end{equation}
with all the remaining quantities kept fixed. 

This transformation
switches between systems with all the constants \(\gamma_q\)
transformed according to the parity of \(q\): the constants remain
identical
for even \(q\)'s and change sign for odd \(q\)'s; in particular
\(\gamma'_1=-\gamma_1\) and \(\gamma'_2=\gamma_2\). At the same time
the order parameter \(\psi_n\) is transformed by the
sign-alternating factor; in particular the states of uniform
order parameter are
transformed into alternating ones, and {\em vice versa}.

In consequence the energy of the Josephson's coupling
is invariant with respect to transformation (\ref{trans}).
This invariance extends to the Josephson part of the total
free energy functional (\ref{fj}).
Two hybrid systems related by the transformation (\ref{trans}) have
identical ground state energies, excitation energy spectra, and all
the thermodynamical properties, including \(T_c\). The shapes of the
corresponding ground states are, however, substantially different.

   The usefulness of the above transformations stems
from the fact that they allow to obtain a solution valid for one
system from a solution valid for another -- anisotropically similar --
system. Analogous transformations, known as anisotropic scaling,
have been defined in reference \cite {Blatter+ 94}. However,
although very useful as a tool for getting approximate estimations,
they do not imply exact invariance. In opposition to that, our
anisotropic similarity transformations are exact. In consequence,
transformed
exact solutions for the depart system are always guaranteed to be
also exact solutions for the target system.



\thebibliography{12}
\bibliography{}

\end{document}